\documentclass[twocolumn,longbib]{aastex701}

\usepackage{natbib}
\usepackage{amssymb}
\usepackage{color}
\usepackage{soul}

\submitjournal{PSJ}

\begin{document}

\title{A change of the rotation period of asteroid (65803)~Didymos caused by the DART impact}

\author[0000-0003-4914-3646]{Josef \v{D}urech}
\affiliation{Charles University, Faculty of Mathematics and Physics, Institute of Astronomy, V Hole\v{s}ovi\v{c}k\'ach 2, 180\,00 Prague, Czech Republic}
\email{durech@sirrah.troja.mff.cuni.cz}

\author[0000-0001-8434-9776]{Petr Pravec}
\affiliation{Astronomical Institute, Academy of Sciences of the Czech Republic, Fri\v{c}ova~298, 251\,65 Ond\v{r}ejov, Czech Republic}
\email{petr.pravec@asu.cas.cz}

\author[0000-0002-1821-5689]{Masatoshi Hirabayashi}
\affiliation{Georgia Institute of Technology, Daniel Guggenheim School of Aerospace Engineering, School of Earth and Atmospheric Sciences, Atlanta, GA 30332, United States}
\email{thirabayashi@gatech.edu}

\author[0000-0002-0054-6850]{Derek C. Richardson}
\affiliation{Department of Astronomy, University of Maryland, College Park, MD 20742}
\email{dcr@UMD.EDU}

\author[0000-0002-3544-298X]{Harrison Agrusa}
\affiliation{Universit\'e C\^ote d'Azur, Observatoire de la C\^ote d'Azur, CNRS, Laboratoire Lagrange, Nice, France}
\email{hagrusa@oca.eu}

\author[0000-0002-9840-2416]{Ryota Nakano}
\affiliation{Georgia Institute of Technology, Daniel Guggenheim School of Aerospace Engineering, Atlanta, GA 30332, United States}
\email{rnakano@gatech.edu}

\begin{abstract}
\noindent
    On 26 September 2022, the NASA Double Asteroid Redirection Test (DART) spacecraft impacted Dimorphos, the secondary component of the binary asteroid (65803)~Didymos. This experiment tested the Kinetic Impactor technology for diverting dangerous asteroids. Due to the impact, the binary system's angular momentum has changed, resulting in a significant change in the orbital period of Dimorphos. Precise values of the pre- and post-impact orbital periods were derived from a large set of photometric light curves measured for the Didymos-Dimorphos system during six apparitions from 2003 to 2023. We used these data to detect a possible change in the rotation period of the primary as a consequence of the impact. We analyzed the binary system's light curves using the binary asteroid light curve decomposition method. We selected parts of the light curves covering orbital phases outside mutual events, which represent the primary rotational light curves. We applied the light curve inversion method to construct a convex shape model of Didymos and determine its rotation period before and after the impact. These two periods were treated as independent free parameters of the modeling. We found a value of $2.260\,389\,1 \pm 0.000\,000\,2$\,h for the pre-impact period and $2.260\,440 \pm 0.000\,008$\,h for the post-impact period. Their difference $0.18 \pm 0.03$\,s is small yet significant, indicating that the rotation of Didymos became slower after the DART impact. The most plausible physical explanation is Didymos's post-impact reshaping, making its shape more oblate. 
\end{abstract}


\section{Introduction}
    The main goal of NASA's Double Asteroid Redirection Test (DART) mission was to test the Kinetic Impactor technology for diverting hazardous asteroids by impacting Dimorphos, the secondary component of the binary asteroid (65803)~Didymos \citep{Dal.ea:23}. The impact occurred on 26 September 2022. According to the predictions, the mutual orbital period changed because of the momentum transfer from the impactor to the binary system. The measurement of this period change was necessary to evaluate the efficiency of the momentum transfer, which was a critical yet unknown parameter of the experiment. The orbital period was measured precisely from photometric observations of mutual events (eclipses and occultations) between Didymos and Dimorphos \citep{Pra.ea:22, Mos.ea:24}. The data set covered six apparitions from 2003 to 2023. Analysis of this data and independent radar observations showed that the Dimorphos orbital period decreased from the pre-impact value of $\sim 11.92$\,h to the post-impact value of $\sim 11.37$\,h, which is a difference of about 33 minutes \citep{Tho.ea:23}. Further photometric observations and their analyses have led to a refined estimate of the orbital period change of $-33.240 \pm 0.024$\,min \citep{Sch.ea:24, Nai.ea:24}.

    Although the main effort in the pre-impact analyses, including modeling and interpretations of the DART images and ground-based observations of Didymos and Dimorphos, was devoted to the aspects related to the mutual orbit, ejecta modeling, and shape modification of Dimorphos, there were studies suggesting that the primary Didymos may have been affected by the impact, too \citep{Hir.ea:17, Hir.ea:19, Riv.ea:21, Ric.ea:22, Hir.ea:22, Nak.ea:22}. However, given the limited observational data, whether the reshaping-driven spin change indeed occurred has not been confirmed \citep{Dal.ea:23,Ric.ea:24,Nak.ea:2024}. These studies motivated our work to investigate a possible change in the rotation period of Didymos, or at least to place an upper limit on such a change, which is essential for theoretical models of the post-impact evolution of the Didymos--Dimorphos system. 

    The original estimate of the rotational period of Didymos came from \cite{Pra.ea:06}, who measured a synodic value of 2.2593\,h from their photometric observations taken in November--December 2003 and estimated its synodic-sidereal period difference as $\pm 0.0008$\,h. \cite{Nai.ea:20} corrected the synodic period estimate assuming that the rotational axis of Didymos was aligned with the normal to the orbit plane of Dimorphos around Didymos and obtained a sidereal period estimate of $2.2600 \pm 0.0001$\,h. In this work, we refine the pre-impact Didymos rotation period estimate by nearly three orders of magnitude by modeling the complete pre-impact photometry data taken for Didymos-Dimorphos from 2003 until the DART impact on 26 September 2022 and determine its post-impact rotation period from observations taken from October 2022 to February 2023.
    
\section{Light curve dataset}
\label{sec:lightcurves}
    In this study, we applied the photometric data set used to detect the change in the orbital period of Dimorphos \citep{Pra.ea:22, Mos.ea:24}, but we aimed to detect the change in the primary's rotation period. The complete set covers six apparitions over twenty years. The pre-impact data were collected from 20 November 2003 to 26 September 2022. The post-impact observations we used cover four months, from 24 October 2022 to 21 February 2023. Although the Didymos--Dimorphos system was photometrically observed immediately after the DART impact, we did not use the data from the first four weeks after the impact because the data were contaminated by the light scattered by the dust ejected by the impact; the contamination decreased to a negligible level for our modeling by 2022 October 24. The primary purpose of the Didymos--Dimorphos photometric observations campaign was to derive the orbital parameters of Dimorphos, accurately determine the orbital period, and detect its change after the impact. To do so, it was necessary to observe mutual events in the system that define the geometry of the orbit and the orbital period. 

    However, the aim of our work was different -- we wanted to determine the primary's rotation period. Therefore, we used only a subset of the photometric measurements, only those taken at orbital phases outside the mutual events between Didymos and Dimorphos.
    We used the outside-event data that were derived in  \cite{Pra.ea:06} and \cite{Pra.ea:22} by applying the binary asteroid light curve decomposition method for the pre-impact data, and we applied the same light curve decomposition method as described in \cite{Pra.ea:22} to get the outside-event data from the post-impact observations.
    Because the binary system was not resolved, the observed light curves corresponded to a combined signal from the primary and the secondary. We neglected any shape effects of the secondary and assumed that its contribution to the total light flux from the binary system was constant. This assumption was held pretty well before the impact, as the original shape of Dimorphos was very close to an oblate spheroid \citep{Dal.ea:23}. It was also a reasonable approximation after the impact, as the rotational lightcurves of Dimorphos observed after the impact had low amplitudes from 0.008 to 0.031\,mag \citep{Pra.ea:24}.) 
    We assumed the Dimorphos-Didymos mean-diameter ratio of 0.20 \citep[]{Sch.ea:24}. We assumed that the signal from the secondary was proportional to its projected area; thus, for each light curve, we computed its mean flux and subtracted 4\% of the mean signal for all photometric points. This correction resulted in a slight increase in the observed amplitude of the primary light curve. 
    A subtraction of the actual rotational light curve of the secondary, which was in an excited, non-principal axis (NPA) spin state after the impact, was not feasible as its actual NPA rotation was not uniquely determined and it behaved quasi-chaotically due to the spin-orbit interaction in the binary system \citep{Pra.ea:24}.  However, as the periods associated with the NPA rotation of the secondary were much longer than the spin period of the primary and its amplitude was low, we found that the presence of the secondary rotational light curve in the data had a negligible effect on the modeling of the rotation of the primary.
    
    Some of the light curves in the 2003--2023 data set were precisely calibrated in standard photometric systems, but most were not, or the accuracy of their absolute calibrations was not sufficient for the purpose of our primary rotation modeling, so we used all the data as relative. We applied the convex light curve inversion method \citep{Kaa.ea:01} to derive a convex shape model of Didymos with the sidereal rotation period as one of the model parameters. We weighted the data by their photometric accuracy, which varied over apparitions and observatories. We estimated the accuracy by fitting a Fourier series to each light curve and assuming that the accuracy corresponds to the root mean square (RMS) of residuals. We put a lower limit of 0.005\,mag on residuals to avoid overweighting data with formal RMS residuals below 0.005\,mag. 

    To estimate the uncertainty of the model parameters described in the next section, we created three times one thousand bootstrap samples of the data. They were generated by randomly selecting the same number of light curves from the original full data set (Sect.~\ref{sec:all_data}) and from original pre- and post-impact data separately (Sect.~\ref{sec:separate_data}). For each light curve, individual points were also chosen randomly.
    
\section{Determination of the rotation period and its change}
    To determine the sidereal rotation period, we used the light curve inversion method of \cite{Kaa.ea:01}, which constructs a physical model of an asteroid by fitting its observed light curves. The shape model is convex, represented by a polyhedron. The rotation state is described by three parameters: the direction of the spin axis in the ecliptic coordinates $\lambda$ and $\beta$, and the rotation period $P$. The advantage of this approach, when $P$ is optimized together with spin and shape parameters, is that having a physical model of the asteroid allows for the sidereal period to be determined with very precise accuracy. The precision depends mainly on the observational timescale -- the longer the interval covered by observations, the better the period precision. If the solution for the sidereal period is unique, that is, if we can unambiguously determine the number of revolutions between the first and the last observation, then the rotation phase is confined to a small fraction of the rotation, typically $\lesssim 10^\circ$. If observations cover an interval $T$ and the rotation period is $P$, then there are $N = T/P$ revolutions within that interval. If we take a ten-degree accuracy in rotation phase as a very conservative upper limit, then in the case of Didymos with the rotation period of about 2.26\,h and 20 years of data, the number of revolutions is $\sim 78,000$, so the relative precision is $3.6 \times 10^{-7}$, which corresponds to $8 \times 10^{-7}$\,h, or equivalently $0.003$\,s. Using the same math for four months of post-impact data leads to a conservative estimate of the period precision of $\sim 0.2$\,s. So, if there was a change in the rotation period on the order of tenths of a second due to the DART impact, we should be able to detect it.
    
    \subsection{Modeling all data together}
    \label{sec:all_data}
        In the first step, we used the whole data set as input for the light curve inversion \citep{Kaa.Tor:01, Kaa.ea:01}.  We assumed that the spin axis direction was the same as the pole of the orbit of Dimorphos, so we fixed it at the ecliptic longitude $\lambda = 310.0^\circ$ and latitude $\beta = -80.4^\circ$, the values determined by \citep{Sch.ea:24}. For the light-scattering model, we adopted the approach of \cite{Kaa.ea:01} and used a combination of Lommel-Seeliger and Lambert models. In this model, the phase-angle dependence is described by a function composed of linear and exponential terms with three parameters. Because our data set was relative, that is, the magnitude shifts between individual nights were arbitrary, the part describing the phase function was not needed. 
        
        The best-fit model provided a good fit to the observed light curves; the convex shape model qualitatively agreed with that constructed from radar observations \citep{Nai.ea:20} and, similarly to the radar model, was less oblate than what the DART images of Didymos have shown \citep{Dal.ea:23}. The corresponding rotation period was $P = 2.260389$\,h. We note that convex models generally cannot describe nonconvex features of real asteroid shapes, and their elongation along the rotation axis is often loosely constrained. However, this does not systematically affect their ability to fit disk-integrated light curves and precisely determine the sidereal rotation period.

\begin{figure}[t]
            \includegraphics[width=\columnwidth]{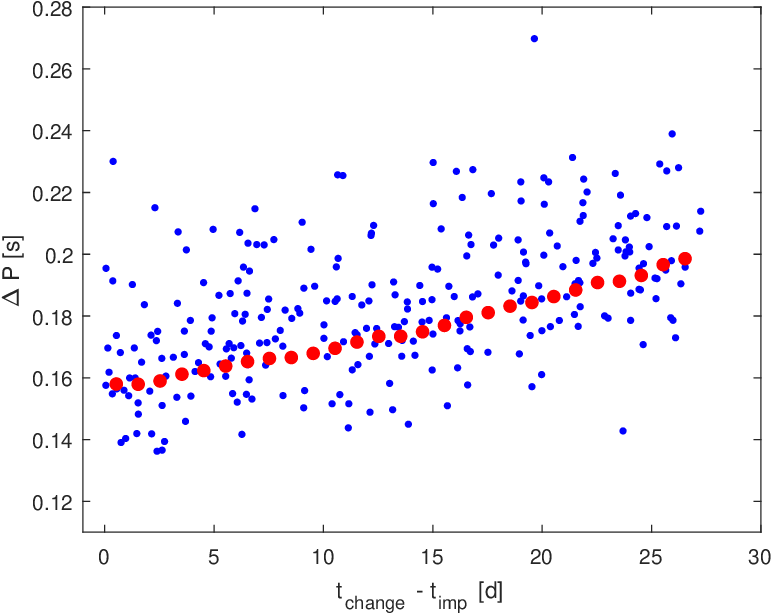}
            \caption{The correlation between $\Delta P$ and $t_\mathrm{change}$. The red points represent the results based on the original data set. The blue points are results based on bootstrap data.}
            \label{fig:t_change}
        \end{figure}

        Then, we introduced another free parameter to the model -- a change $\Delta\omega$ of the rotation rate $\omega$ at a time $t_\mathrm{change}$. We assumed that before $t_\mathrm{change}$, the model rotated with the angular frequency $\omega_1$, and then, after $t_\mathrm{change}$ but before the first post-impact observations, the frequency changed to $\omega_2 = \omega_1 + \Delta\omega$. The rotation phase $\varphi$ of Didymos at some epoch $t$ is then described as
        \begin{equation}
            \varphi(t) = \varphi_0 + \omega_1 (t - t_0) + \Delta\omega (t - t_\mathrm{change})\,,
        \end{equation}
        where $\varphi_0$ is some initial rotation phase at the time $t_0$, $\Delta\omega$ is zero for $t < t_\mathrm{change}$ and nonzero for $t > t_\mathrm{change}$. This model assumes that the rotation period changed instantaneously at the time $t_\mathrm{change}$ from $P_1 = 2\pi / \omega_1$ to $P_2 = 2\pi / \omega_2 = 2\pi / (\omega_1 + \Delta\omega)$. However, the parameter of interest, which is the difference between periods $\Delta P \equiv P_2 - P_1$, is correlated with the unknown $t_\mathrm{change}$. In Fig.~\ref{fig:t_change}, we show how $\Delta P$ depends on $t_\mathrm{change}$. The modelled $\Delta P$ varies from 0.16\,s when $t_\mathrm{change}$ coincides with the time of impact $t_\mathrm{imp}$, up to 0.20\,s when $t_\mathrm{change}$ is put at the beginning of the post-impact data. All combinations of $\Delta P$ and $t_\mathrm{change}$ provide practically the same fit of the data. If we put $t_\mathrm{change}$ close to $t_\mathrm{imp}$, the required $\Delta P$ is smaller because the phase shift with respect to pre-impact rotation accumulates over the whole interval of 27 days, not covered by data. On the other hand, putting $t_\mathrm{change}$ just before the first post-impact light curve requires larger values of $\Delta P$ to fit the phase shift of post-impact data. Although it is more physically plausible to expect the period to change just after the impact rather than at the end of the gap in the data, the model does not constrain this.

\begin{figure}[t]
            \includegraphics[width=\columnwidth]{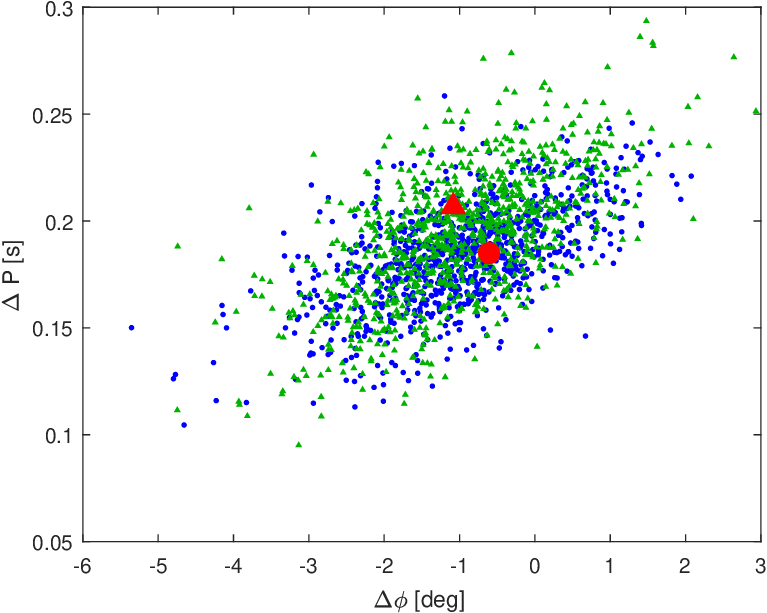}
            \caption{The dependence between $\Delta P$ and $\Delta\varphi$. The red circle represents the best-fit values for the original data set, with the spin pole of Didymos fixed in the direction of the orbital pole of Dimorphos. Blue points are based on bootstrap samples. When the spin axis direction is optimized, the best-fit solution for $\Delta P$ and $\Delta\varphi$ moves to values represented by the red triangle. Bootstrap samples with free pole direction are shown as green triangles.}
            \label{fig:shift}
        \end{figure}

        To avoid the dependence of $\Delta P$ on $t_\mathrm{change}$, we applied a more general model. We assumed that the angular frequency was constant before the impact $t_\mathrm{imp}$ and after the beginning of the post-impact data $t_\mathrm{post}$. Any variations of the angular frequency occurred during the interval that is not covered by photometry. During this interval, the rotation phase increased from $\varphi_\mathrm{imp}$ to $\varphi_\mathrm{post}$ according to
        \begin{equation}
            \int_{t_\mathrm{imp}}^{t_\mathrm{post}} \left(\omega_1 + \Delta\omega(t)\right) \, \mathrm{d}t = \int_{t_\mathrm{imp}}^{t_\mathrm{post}} \mathrm{d}\varphi = \varphi_\mathrm{post} - \varphi_\mathrm{imp}\,,
        \end{equation}
        where $\Delta\omega(t)$ describes some general differential evolution of rotation with respect to $\omega_1$. In the previous model, $\Delta\omega(t)$ was a step function with step size $\Delta\omega$ at time $t_\mathrm{change}$. The general model now introduces another free parameter $\Delta\varphi$ that describes the phase shift due to $\Delta\omega(t)$, that is 
        \begin{equation}
            \Delta\varphi = \int_{t_\mathrm{imp}}^{t_\mathrm{post}} \Delta\omega(t)\, \mathrm{d}t\,.
        \end{equation}
        The rotation phase of Didymos is described as
        \begin{eqnarray}
            \varphi(t < t_\mathrm{imp}) & = &\varphi_0 + \omega_1 (t - t_0)\,, \\
            \varphi(t > t_\mathrm{imp}) & = &\varphi_0 + \omega_1 (t_\mathrm{post} - t_0) + \omega_2 (t - t_\mathrm{post}) + \Delta\varphi\,, \nonumber
        \end{eqnarray}
        where $\omega_1$, $\omega_2$, and $\Delta\varphi$ are free parameters of optimization. Fig.~\ref{fig:shift} shows the result for the original data set and bootstrap samples. The period change $\Delta P$ is still correlated with $\Delta\varphi$, but now the value of $\Delta\varphi$ is the result of optimization, not our choice.

        We found the best-fit parameters to be $P_1 = 2.260389$\,h and $P_2 = 2.260440$\,h, which means that the period has increased by $\Delta P = 0.000051\,\text{h} = 0.18\,\text{s}$. The improvement of this changing-period fit with respect to a model with an unperturbed constant rotation rate $\omega$ was small but statistically significant; the $\chi^2$ has decreased by about $4\%$ with the degrees of freedom of about 18,000. We also tested the null hypothesis that the residuals of the two models have the same variance (against an alternative hypothesis that the residuals of the changing-period model are smaller). The F-test yielded a probability of 2.3\%. We estimated the uncertainty of $\Delta P$ by repeating the inversion for a thousand bootstrap samples of photometric data (see Sect.~\ref{sec:lightcurves}). The standard deviation of $\Delta P$ was 0.02\,s. The value and uncertainty of $\Delta P$ depend only weakly on our assumption of fixed spin pole direction. If we let parameters $\lambda$ and $\beta$ be optimized, the best-fit parameters are: $\lambda = 317^\circ$, $\beta = -77^\circ$, $\Delta\varphi = -1.1^\circ$, and $\Delta P = 0.21$\,s. The mean values and standard deviations from bootstrap samples are: $\lambda = (313 \pm 4)^\circ$, $\beta = (-77 \pm 2)^\circ$, and $\Delta P = 0.19 \pm 0.03$\,s. 
        
\begin{figure}[t]
            \includegraphics[width=\columnwidth]{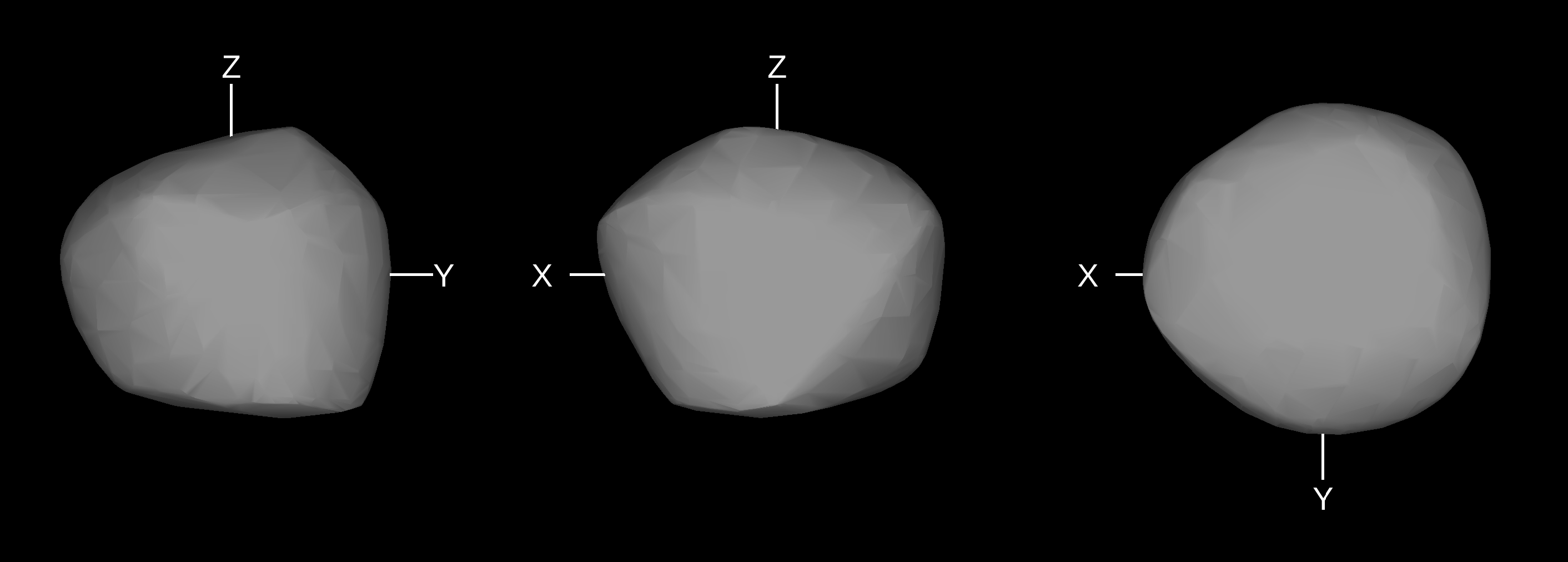}
            \caption{Convex shape model of Didymos reconstructed from its light curves.}
            \label{fig:shape}
        \end{figure}

\begin{figure*}[t]
            \plotone{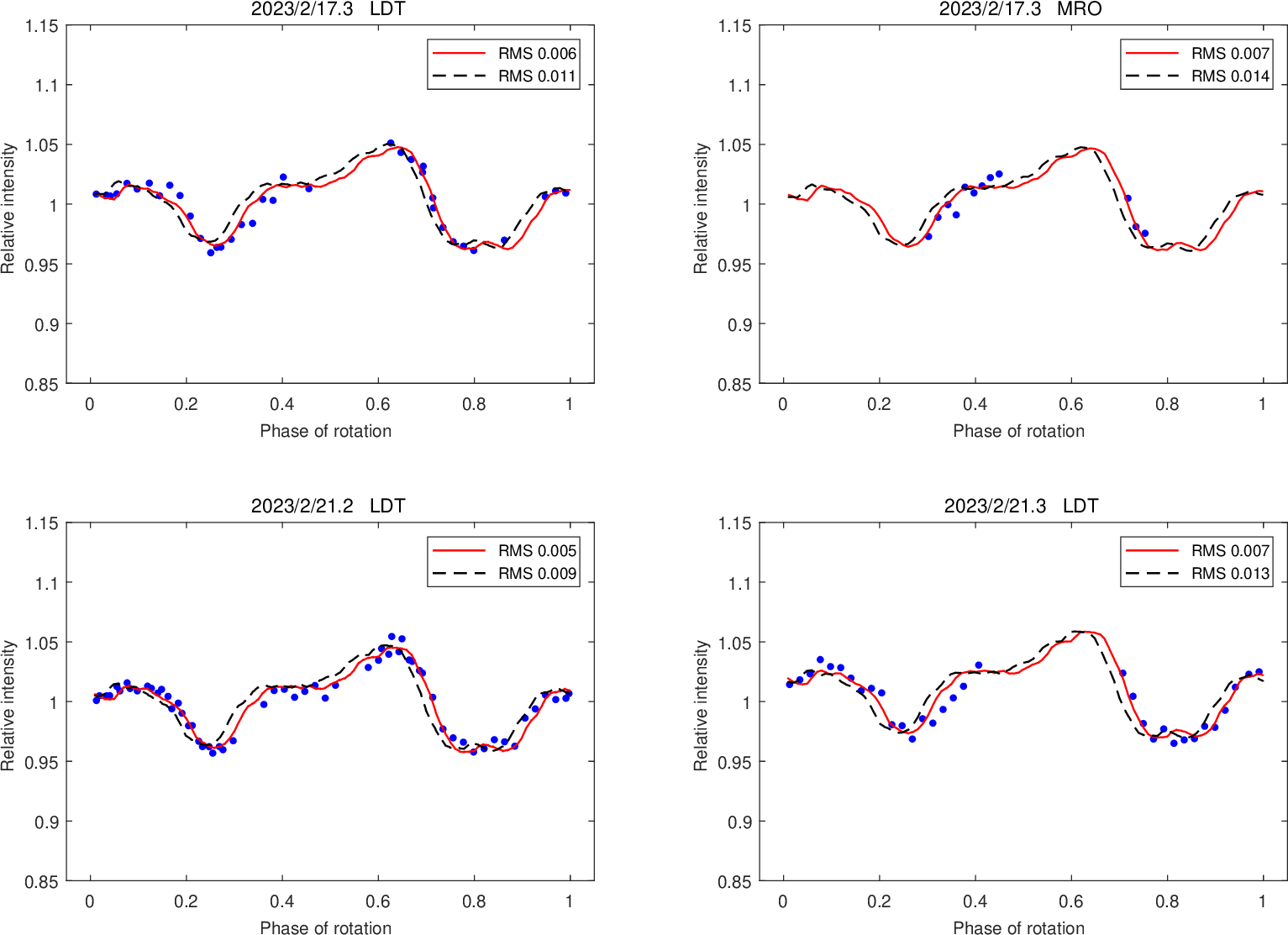}
            \caption{Example light curves at which the difference between the constant-period model (dashed black curve) and the model with the two different periods (solid red curve) is most prominent. The blue points are measured photometric data normalized to the unit mean brightness.}
            \label{fig:lcfit}
        \end{figure*}

        The convex shape model corresponding to the best-fit solution is shown in Fig.~\ref{fig:shape}. It qualitatively agrees with that derived from radar observations by \cite{Nai.ea:20}. For most of the data, there is practically no difference between the constant $\omega$ model and the model with $\omega_1 \neq \omega_2$; the fit is visually almost the same. The only visible difference is apparent in the last couple of light curves, as shown in Fig.~\ref{fig:lcfit}. All four light curves exhibit the same trend, with the constant-period model lagging behind the observed data. Three of these light curves were taken with the Lowell Discovery Telescope (LDT), and one was taken at the Magdalena Ridge Observatory (MRO); therefore, it is unlikely that a systematic timing error in the observations caused the shift. 
        
    \subsection{Modeling pre- and post-impact data separately}
    \label{sec:separate_data}
        The detected change of 0.18\,s in the rotation period described above is statistically significant. However, we wanted to check that it might not be affected by some possible hidden systematic errors in the data. To at least partly verify that this was not the case and estimate the uncertainties of the periods, we ran the light curve inversion for one thousand bootstrap samples of pre- and post-impact observations. As in the case of the complete data set, the spin axis direction was fixed on the values of the orbital pole, so the only free parameters were the rotation period and shape parameters. 
        
        Apart from splitting the data into pre- and post-impact subsets, we also created two subsets from data covering 2017--2021 and 2020--2022 (pre-impact only) apparitions. We used them to check the stability of the pre-impact solution.

\begin{figure}[t]
        \includegraphics[width=\columnwidth]{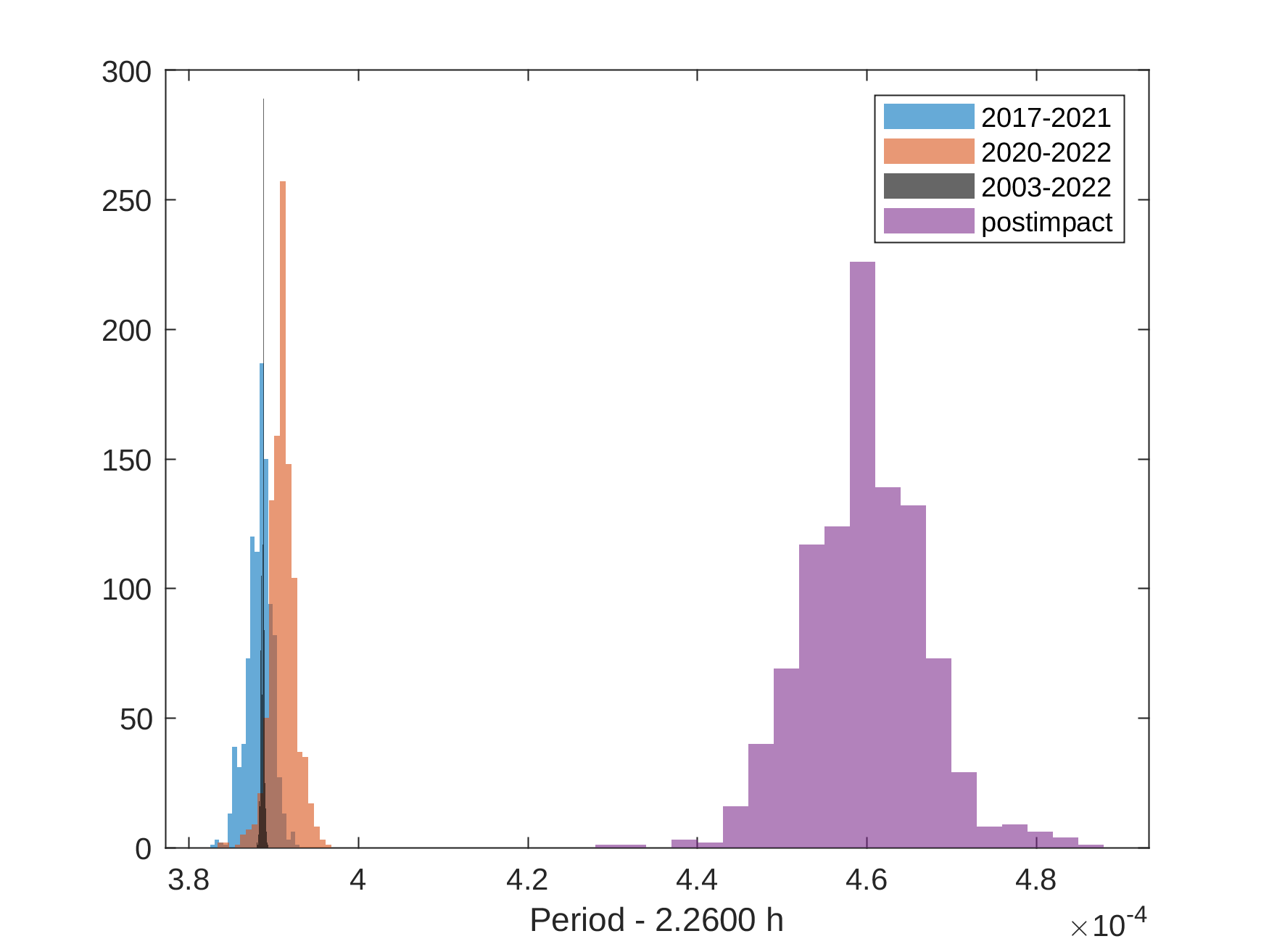}
        \caption{Histograms of periods determined from bootstrap samples for four subsets of light curve data. The post-impact periods are separated from the pre-impact periods.}
        \label{fig:histograms}
    \end{figure}

        The distribution of sidereal rotation periods obtained from bootstrap data is shown in Fig.~\ref{fig:histograms}. The main results visible in these histograms are: (i) the post-impact periods are consistently longer than pre-impact periods, and (ii) all three pre-impact data sets have the same rotation periods. The spread of the distributions is our estimate of the uncertainty of the periods. The standard deviation of pre-impact bootstrap periods is $1.71 \times 10^{-7}$\,h, which is 0.00061\,s. For the 2017--2021 light curves, it is $1.13 \times 10^{-6}\,\text{h} = 0.0041\,\text{s}$; for the 2020-2022 light curves, it is  $1.81 \times 10^{-6}\,\text{h} = 0.0065\,\text{s}$. Thus, all three pre-impact subsets exhibit the same rotation period within the uncertainty, resulting in their histograms overlapping. As expected, the dispersion of the periods is related to the interval of observations; the longer the interval, the smaller the dispersion. The uncertainty of the period derived from the bootstrap distribution is about five times smaller than our conservative estimate from Sect.~\ref{sec:lightcurves} based on the 10-deg shift. 
        
        The distribution of post-impact periods is shifted towards longer periods with the mean value of 2.260462\,h (corresponding to $\Delta P = 0.26$\,s) and the standard deviation of $7.7 \times 10^{-6}\,\text{h} = 0.028\,\text{s}$ (seven times smaller than the conservative estimate). The post-impact period derived from the full-dataset model was 2.260440\,h, which is slightly different but still consistent (within $3\sigma$ uncertainty) with the bootstrap distribution. For comparison, the rotation period derived from the original post-impact light curve data set was 2.260457\,h, corresponding to $\Delta P = 0.24$\,s, also larger than the value derived in Sect.~\ref{sec:all_data}. This difference likely stems from the different modeling approach -- while in Sect.~\ref{sec:all_data} the shape model was the same for pre- and post-impact data (Fig.~\ref{fig:shape}), here the two data sets are treated as fully independent with independent shape models. 
        
        If we take the standard deviation as an estimate for the $1\sigma$ uncertainty of the period determination, then the detected shift in the period from Sect.~\ref{sec:all_data} is $0.18 \pm 0.03$\,s. It is formally a $6\sigma$ detection.

    \subsection{Possible systematic effects}    
        The bootstrap approach described above did not reveal any systematic effect in the data -- pre-impact data have a different rotation period than post-impact data for all bootstrap samples. However, some systematic model errors might affect the period determination.

        One of the simplifications of the model was that a convex polyhedron represented the shape, whereas the actual body of Didymos is nonconvex with complicated surface features. We aimed to determine precisely the rotation period; the convex shape model was just a by-product of the inversion, so an imprecise shape model was not an obstacle. Even if shape nonconvexities could deform and shift the light curves of a couple of degrees in phase -- which was what we detected and interpreted as a period change -- such shifts would depend on the viewing and illumination geometry, which was changing, so there should not be any systematic net effect. We created synthetic light curves with a nonconvex body to test this argumentation and simulated the inversion. We used a nonconvex shape of asteroid (3)~Juno \citep{Vii.ea:15b} -- not very elongated asteroid with moderate nonconvex features, Hapke's scattering model with the scattering parameters of an average S-type asteroid \citep{Li.ea:15}, and the same pole direction as for Didymos with the rotation period of 2.26039\,h. The testing light curves were generated for the same observing geometry as the actual data, and we added a Gaussian noise of 2\%. We performed bootstrap resampling the same way as for real data, ran the light curve convex inversion, and computed the dispersion of the period values. The standard deviation of post-impact periods was $7.07 \times 10^{-6}\,\text{h} = 0.026\,\text{s}$, which was about the same as with the real Didymos data. Pre-impact test light curves provide periods with the standard deviation of $6.10 \times 10^{-8}\,\text{h} = 0.00022\,\text{s}$, which was three times lower than with the real data, likely because the noise of some of the pre-impact observations was larger than 2\% (see \citep{Pra.ea:22}). There was no significant difference between the mean bootstrap period and the original correct value. This exercise demonstrated that the period determination and its uncertainty are not sensitive to the convex shape approximation and the scattering model.

        Another assumption of our model was that the spin axis of Didymos was parallel with the orbital pole direction; that is, there was no inclination between the orbital plane of Dimorphos and the equator of Didymos. This is a reasonable assumption supported by the fact that no nodal precession of the orbit was observed \citep{Sch.ea:24, Nai.ea:24}.  Nodal precession would be inevitable in the case of a nonzero inclination. We checked the robustness of our results concerning the spin pole direction. We repeated the bootstrap modeling with spin axis parameters $\lambda$ and $\beta$ being subjects of optimization. This had only a little effect on the results. The standard deviation of the pre-impact period increased slightly to $1.74 \times 10^{-7}$\,h, for post-impact data to $10.0 \times 10^{-6}$\,h. The results were qualitatively the same because, even if the pole was allowed to be free, the parameters $\lambda$ and $\beta$ converged to values close (within a few degrees) to the original ones. The dispersion across different bootstrap realizations was approximately two degrees.
        
        To further test the sensitivity of the model to the pole direction, we used $3\sigma$ uncertainties of the orbital pole reported by \cite{Sch.ea:24}, which were $\pm 15^\circ$ in $\lambda$ and $\pm 1.9^\circ$ in $\beta$. Using four combinations of the pole position at the boundary of the $3\sigma$ interval, we obtained $\Delta P$ in the range 0.14--0.21\,s.
        
        We also checked our assumption that the rotation period was constant during the intervals covered by the data. We employed the same approach as \cite{Kaa.ea:07} or \cite{Dur.ea:22} to model secular changes in the rotation period, introducing an additional model parameter $\mathrm{d}\omega / \mathrm{d} t$. In single asteroids, this parameter can be interpreted as the evolution of the rotation rate due to the thermal YORP effect \citep{Vok.ea:15}. In our context, we used it as just another degree of freedom of the model. For pre- and post-impact data modeled separately, this new parameter did not affect the solution; the fit did not improve, and the residuals remained practically the same. We detected no secular changes, so our constant period assumption was acceptable.

        As described in Sect.~\ref{sec:lightcurves}, both binary system components contribute to the measured light curves. We subtracted the mean signal from the secondary, assuming that its contribution to the total light flux from the binary system was constant. However, the actual post-impact shape of Dimorphos is not a spheroid (see Sect. 2), so this is another simplification of our model. Moreover, according to \cite{Sch.ea:24}, the pre-impact size ratio of 0.22 changed after the impact to 0.20, indicating that Dimorphos's size or shape has changed slightly due to the impact. Moreover, the post-impact signal from the secondary is not strictly periodic because, according to \cite{Pra.ea:24}, it is likely that Dimorphos was excited into a non-principal-axis rotation by the impact. Fortunately, all these unmodeled effects regarding the secondary's shape, size, and rotation, and their contribution to the signal are ``orthogonal'' to the trend we are looking for. These effects do not modify the period of the Didymos rotational light curve. This is consistent with dynamical expectation: the almost perfect oblate shape of Didymos resists any significant torque from changing the rotational period \citep{Hir.ea:19}. We obtained the same result for $P_1$ and $P_2$ even when we did not subtract any secondary signal, which showed us that the exact value of the subtracted signal was not important.

\section{Possible physical mechanism/explanation}

    Although a detailed analysis is beyond the scope of this work, the most likely explanation of the observed slowdown of Didymos's rotation is a reshaping of the body, which became more oblate than its pre-impact shape due to a small-scale landslide liberated by the accretion of ejecta. Below, we briefly discuss this mechanism. 
    
    Didymos’s pre-impact spin period was ${\sim}2.260$\,h, which is close to the spin limit for S-type asteroids of similar sizes. Early radar measurements suggested that Didymos had a top shape \citep{Nai.ea:20}, though DART's observations revealed its shape to be much more oblate \citep{Dal.ea:23, Bar.ea:24}. Previous studies using radar-derived shape model predicted that Didymos was close to its structural failure limit, where additional spinup would lead to irreversible deformation modes, including surface mass movement and/or internal failure \citep{Hir.ea:17, Hir.ea:19, Zha.ea:21, Hir.ea:22, Ric.ea:22}. Based on Didymos's updated shape model and bulk density of ${\sim}2800\,\mathrm{kg}\,\mathrm{m}^{-3}$, some small cohesive strength, less than $\sim1$\,Pa, is required to resist surface failure \citep{Bar.ea:24}. 

    Owing to Didymos's fast spin, any internal or surface failure may cause material to move outwards, away from the spin axis. Earlier works predicted that Didymos's rotation rate would decrease to compensate for its increased moment of inertia \citep{Hir.ea:17, Hir.ea:19}. Based on the early radar shape model, and assuming angular momentum and volume are conserved and that any shape change occurs uniformly, a decrease in the length of Didymos's spin axis by $\sim1$\,cm would have an associated $\sim0.1$\,s increase in the spin period \citep{Nak.ea:22}. Updating the calculations, we find the measured $\sim0.18$\,s spin period increase should correspond to a $\sim$13\,mm decrease in the spin axis and a $\sim$9\,mm increase in the equatorial axes of the Dynamically Equivalent Equal-Volume Ellipsoid (DEEVE) of the current Didymos shape model. If the slowdown indeed results from reshaping, this small level of change is most likely due to surface mass movements, rather than a large-scale global reshaping event. If the surface mass movements (possible landslides) were localized rather than global, they would be much larger in magnitude than the estimated $\sim$13\,mm change of the Didymos DEEVE. A reasonable explanation is that low-speed bombardments of the DART impact ejecta liberated surface materials, causing mass flows towards lower latitudes and thus increasing the moment of inertia along the short axis. 

    Multiple mechanisms may also change Didymos's spin period, though, to our knowledge, they do not offer a sufficient explanation for this rotational slowdown. First, the accretion of impact ejecta could modify Didymos's spin. However, the DART spacecraft hit Dimorphos in a retrograde sense relative to Didymos' spin pole (and the binary orbit pole), generating predominantly prograde ejecta. Therefore, the accretion of DART ejecta onto Didymos adds angular momentum rather than subtracts it, shortening the spin period. Another mechanism is spin-orbit coupling between the spin of Didymos and the orbit of Dimorphos. However, Didymos's oblate shape and extremely fast rotation compared to Dimorphos's mean motion result in negligible interaction between Didymos's rotation state and the mutual orbit, making this possibility unlikely. In addition, any spin-orbit coupling resulting in a reduction in Didymos' rotation rate would come with an associated increase in the mutual orbit period and semimajor axis by conservation of angular momentum. However, it appears that Dimorphos's orbit period has only shortened since the DART impact \citep{Nai.ea:24,Sch.ea:24}, although this mechanism needs further investigation \citep{Agr.ea:25}. The YORP effect is another mechanism that could change the rotation rate of Didymos. Based on the known YORP strengths for other small asteroids \citep{Durech2024}, it is extremely unlikely that YORP would have measurably changed Didymos's spin rate on such a short timescale.  In summary, the above possibilities are all opposite to the reported observations or unlikely. 
    
\section{Conclusions}

    By performing the light curve inversion of Didymos photometry, we revealed a change in its sidereal rotation period that coincides with the time of the DART impact. 
    The pre-impact period was $P_1 = 2.260\,389\,1 \pm 0.000\,000\,2$\,h and the post-impact one $P_2 = 2.260\,440 \pm 0.000\,008$\,h. This means that the Didymos rotation period increased by $0.18 \pm 0.03$\,s. Our model does not constrain the exact time of this change or its physical cause. All we can say is that it happened -- instantaneously or continuously -- during the first $\sim 4$ weeks after the DART impact that were not covered by the data set we used. 
    
    The most plausible explanation for the asteroid's spin-down is the increase in its moment of inertia along the spin axis. Given its fast rotation and oblate shape, Didymos may have been close to its failure limit in the interior or on the surface before impact \citep{Hir.ea:17, Hir.ea:19}. If any reshaping indeed occurred, Didymos's principal moment of inertia would likely increase, leading to spin-down. The period change of $\sim 0.18$\,s corresponds to extending the Didymos DEEVE equatorial axes by $\sim 9$\,mm and shortening the spin axis by $\sim 13$\,mm. The most reasonable is one or more mild mass movements, rather than internal failure, which can otherwise cause a drastic shape change.  Such localized mass movements (possible landslides) could be of a substantial magnitude, and they may be resolved by Hera when it images Didymos in 2026--2027.

    Regarding the observational detection of the period deceleration, future observations will significantly shrink the uncertainty of the post-impact period determination. Due to the still short time base of only four months of the post-impact data, this is the primary source of uncertainty currently. Future observations can also constrain the possible evolution of the post-impact spin rate. So far, we have not detected any; the 2022--2023 data are entirely consistent with the simplest assumption that the post-impact rotation period was constant on the interval covered by observations.

\begin{acknowledgments}
    This work was supported by the grant 23-04946S of the Czech Science Foundation. P.P. was supported by {\it Praemium Academiae} award (no.\,AP2401) from the Academy of Sciences of the Czech Republic. M.H. acknowledges the support from 23-HERAPSP23/80NSSC24K1434. H.A. was supported by the French government, through the UCA J.E.D.I. Investments in the Future project managed by the National Research Agency (ANR) with the reference number ANR-15-IDEX-01.
\end{acknowledgments}



\bibliographystyle{aasjournalv7}
\bibliography{bibliography_all}

\end{document}